\documentclass[a4paper, oneside, twocolumn, notitlepage, 10pt]{extarticle_ecoc}
\usepackage{ecoc}

\begin{document}
\selectlanguage{english}    % Standard Language

%-------------------------------------------------- Title -----------------------------------------------------%

\title{Efficient Multi-Agent Optimization of Optical Power in S+C+L-Band Systems}%

%------------------------------------------------- Authors-----------------------------------------------------%

\author{
    Junzhe Xiao\textsuperscript{†}, Kaida Chen\textsuperscript{†}, Cong Wang\textsuperscript{†}, Zekun Niu*, Minghui Shi, Yanhan Zhou and Lilin Yi*
}

% \author{
%     Juzhe Xiao, Kaida Chen, Cong Wang, Zekun Niu*, Minghui Shi, Yanhan Zhou and Lilin Yi*
% }

\maketitle                  % Create title and author

%------------------------------------------ Description of Authors ----------------------------------------------%

\begin{strip}
    \begin{author_descr}

        State Key Laboratory of Photonics and Communications, School of Integrated Circuits (School of Information Science and Electronic Engineering), Shanghai Jiao Tong University, Shanghai 200240, China,
        *\textcolor{blue}{\uline{zekunniu@sjtu.edu.cn}}, *\textcolor{blue}{\uline{lilinyi@sjtu.edu.cn}}

    \end{author_descr}
\end{strip}

% \setstretch{1.1}
%-------------------------------------------------- Footnote -------------------------------------------------------%
\renewcommand\footnotemark{}
\renewcommand\footnoterule{}
%\let\thefootnote\relax\footnotetext{text}

%-------------------------------------------------- Abstract ---------------------------------------------------------%

\begin{strip}
    \begin{ecoc_abstract}
        % NOTE: Don't use a blank line here but start abstract right away to avoid an extra line break
        We propose an AI Agent tailored for link power management in multi-band systems. In S+C+L band span-level study, the agent efficiently solves various optimization objectives. In network-wide evaluation, it delivers 689.0 Tbps gain in total allocated traffic with merely 303 average interactions per power profile.©2026 The Author(s) 
    \end{ecoc_abstract}
\end{strip}

%-------------------------------------------------- Introduction Section -------------------------------------------------------%

\section{Introduction}
Multi-band transmission (MBT) offers a potential solution to expand the capacity of optical fiber networks \cite{9076329}. In recent years, the S-band emerged as a viable option to further expand system bandwidth \cite{10530896}. In S+C+L-band systems, signal power evolution is significantly impaired by amplifier gain ripples, noise figure (NF) variations, and inter-channel stimulated Raman scattering (ISRS). As a result, both the quality of transmission (QoT) uniformity and the overall system throughput are severely compromised. Previous studies have investigated power optimization methods to enhance the performance of multi-band transmission systems \cite{10892225,9402935,10214139,9748510,11082171,9492399}. 

\begin{figure*}[b]
    \centering
    \includegraphics[width=\textwidth]{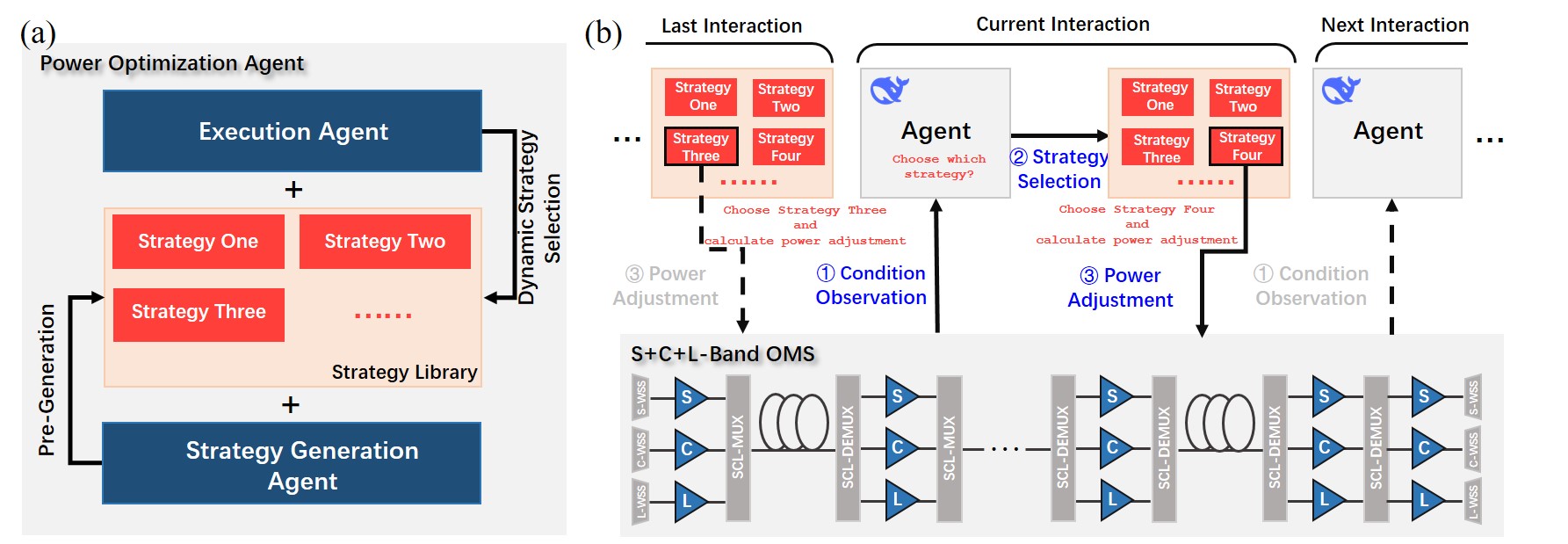}
    \caption{Structure and workflow of the Power Optimization Agent: (a) agent architecture; (b) operational workflow.}
    \label{fig:fig1}
\end{figure*}

Enhancing the capacity of multi-band transmission systems can be modeled as a power optimization problem \cite{7470251}, and power optimization is executed in the model-based or model-free way in previous work. Model-based methods \cite{10892225,10214139,9748510} optimize the power depending on the models for QoT estimation. However, these methods are often limited in achieving optimal power configuration owing to parameter uncertainties and imperfect modeling \cite{10187397}. In contrast, model-free methods \cite{9082752,11082171,9492399} optimize link power via direct interactions with the environment, regarding multi-band transmission systems as black boxes. Nevertheless, such approaches demand a considerable number of iterative interactions with the transmission system to converge \cite{9960788}.

Recently, the rapid advancement of Large Language Models (LLMs) has spurred the emergence of LLM-based AI agents \cite{xi2023risepotentiallargelanguage}. Furthermore, such agents have demonstrated capabilities in solving optimization problems \cite{liu2025agenthpo}, making them promising for multi-band transmission system power optimization.

In this paper, Power-Optimization (PO) Agent is specifically designed to optimize link power in multi-band transmission systems, targeting diverse optimization objectives: (i) maximizing total capacity \cite{9082752}; (ii) flattening the received generalized signal-to-noise ratio (GSNR) \cite{10214139}; (iii) flattening the received optical signal-to-noise ratio (OSNR) \cite{9580565}; and (iv) equalizing the received power \cite{10892225}. During the optimization process, the agent dynamically adopts different optimization strategies based on its interactions with the multi-band system. In the S+C+L-band single span transmission, the proposed agent outperforms the Genetic Algorithm (GA) \cite{9492399} and Particle Swarm Optimization (PSO) \cite{9082752} across all objectives under limited interaction budgets. Furthermore, progressive traffic analysis on the Italy network topology demonstrates that PO agent supports a significant enhancement of 689.0 Tbps in total allocated traffic with higher efficiency.

\section{Link power optimization with AI Agent}
Fig.1(a) depicts the architecture of the proposed PO Agent for optical power optimization. Built on a multi-agent framework \cite{ijcai2024p890}, the system consists of two collaborative sub-agents: the execution agent and the strategy generation agent.

The strategy generation Agent is responsible for constructing and maintaining a strategy library. Each entry in this library is a specialized function that outputs link power adjustments, tailored specifically for optical power optimization. By replacing complex numerical calculations with these pre-defined strategies, PO Agent effectively overcomes the inherent limitations of LLMs in precise numerical computation \cite{li-etal-2025-exposing}.

The execution agent interacts with the multi-band transmission system to adjust link power during the optimization process. First, it observes the current system condition and then selects an appropriate strategy from the pre-defined library based on these observations. Once the strategy calculates the required power adjustment, the system updates the link power accordingly, initiating the next interaction cycle with a new observation.

Fig.1(b) illustrates the detailed operational workflow of the PO Agent. In this framework, the execution agent dynamically retrieves optimal strategies from a library curated by the strategy generation agent. The optimization process continues until the PO Agent terminates autonomously or reaches a pre-set maximum number of interactions.

\begin{figure*}[b]
    \centering
    \includegraphics[width=\textwidth]{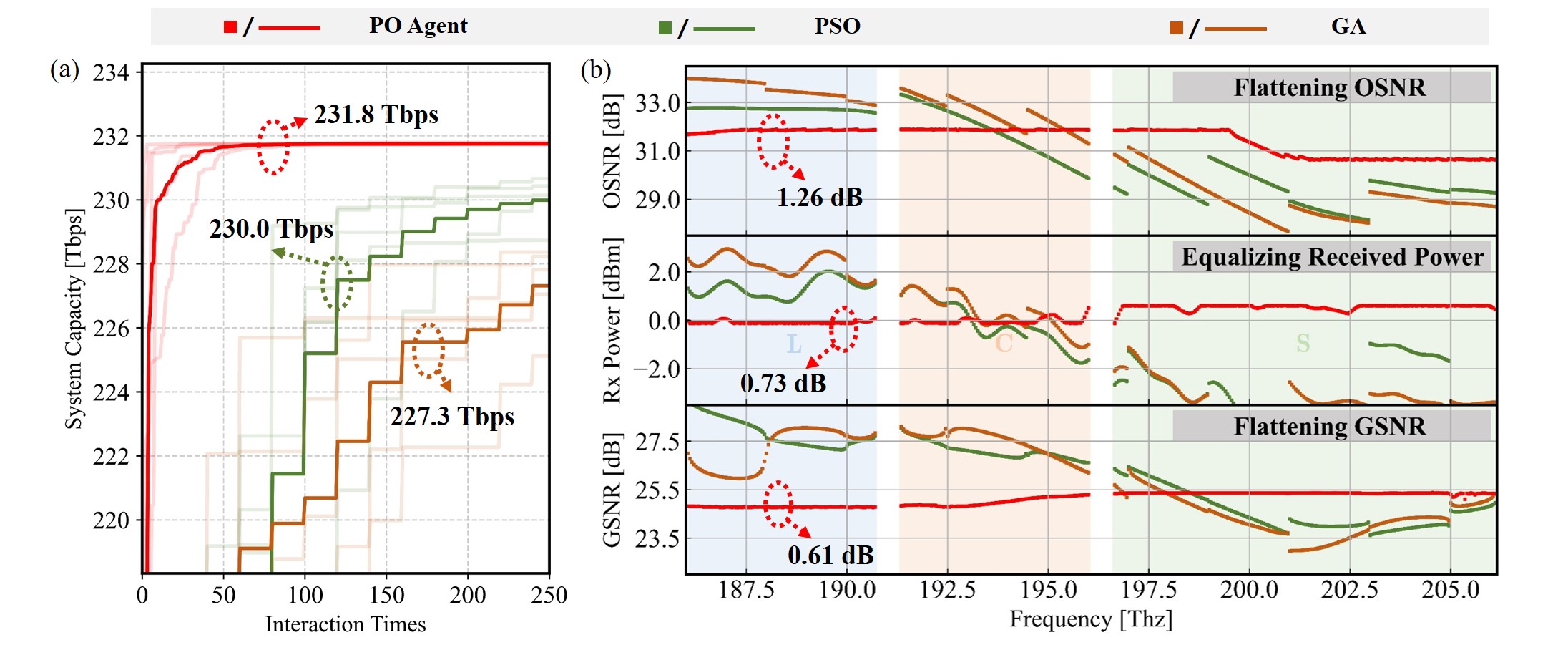}
    \caption{Span-level optimization results: (a) Total capacity as a function of interaction time; (b) Spectral profiles of the received OSNR (top), power (middle), and GSNR (bottom) after optimization.}
    \label{fig:fig2}
\end{figure*}

The objectives of link power optimization are embedded within the PO Agent as system prompts, including maximizing total capacity, flattening the received GSNR and OSNR, and equalizing received power. At the commencement of the optimization, the PO Agent parses the user's requirements and initiates the optimization process. Concurrently, workflow prompts and action rules are embedded within the agent to enhance the reliability of the optimization.

Built upon the LangChain framework \cite{langchain}, the PO Agent employs the ReAct paradigm \cite{yao2023reactsynergizingreasoningacting} and leverages DeepSeek-V3.2 \cite{deepseekai2025deepseekv32pushingfrontieropen} as its underlying language model.

\section{Span-Level Optimization}

In the span-level study, the launch power of each channel is optimized over an 80 km span of ITU-T G.652D standard single-mode fiber (SSMF) transmission, with the aim of achieving four objectives: maximizing total capacity, flattening the received GSNR and OSNR, and equalizing received power. According to the ITU-T G.694.1 standard \cite{Luo:22}, the frequency ranges for these bands are defined as 185.975 THz \textasciitilde 190.775 THz, 191.275 THz \textasciitilde 196.075 THz, and 196.575 THz \textasciitilde 206.175 THz, respectively. Each band operates on the ITU-T 50 GHz WDM grid, with transceivers configured for a symbol rate of 32 GBaud. Consequently, 384 channels (96 channels in L-band, 96 channels in C-band, and 192 channels in S-band) are employed and optimized in total for S+C+L-band transmission in this study. The multi-band transmission is simulated with GNPy \cite{9035688}, which considers the amplifier gain ripples, NF variations, Kerr nonlinearity and ISRS.

Meanwhile, GA \cite{9492399} and PSO \cite{9082752} are employed as baseline algorithms for performance comparison. For both methods, the maximum interaction budget is set to 250 and the optimization process is repeated for 5 times. Additionally, ideal flexible transceivers are assumed, enabling continuous bit-rate adaptation based on the available GSNR \cite{9402935}. To enhance convergence speed of GA and PSO, the adjacent channel power approximation \cite{9082752} is incorporated into the optimization process, with the number of adjacent channels set to 40. On the other hand, PO Agent directly observes and adjusts the launch power of all channels.

Fig. 2(a) illustrates the total capacity of the 80-km S+C+L-band transmission system as a function of interaction time. As the results show, the total capacity increases rapidly within the first 50 interactions for the PO Agent, whereas PSO and GA converge at a slower rate. Benefiting from this rapid convergence, the PO Agent achieves an average maximum capacity of 231.8 Tbps after 250 interactions, compared to 230.0 Tbps and 227.3 Tbps for PSO and GA, respectively. Fig. 2(b) presents the average optimized results for the remaining targets. After optimization by the PO Agent, the average ripples in received OSNR, power, and GSNR are 1.26 dB, 0.73 dB, and 0.61 dB, respectively. In contrast, ripples obtained with the other optimization methods all exceed 4 dB. 

\section{Network-Wide Optimization}

We also compare the PO Agent with PSO and GA over the Italy 
network \cite{10789615}. As shown in Fig. 3(a), the Italy network has 21 nodes and 35 links, with an average nodal degree of 3.3, average distance between nodes of 239 km, and maximum link length of 617 km. We approximate the real topology with SSMF of 80 km per span, and suppose that the wavelength-selective switch (WSS) for all bands is set after every 3 span transmission for power adjustment. Consistent with the span-level study, network transmission utilizes 384 channels across the S, C, and L bands.

\begin{figure*}[t]
    \centering
    \includegraphics[width=\textwidth]{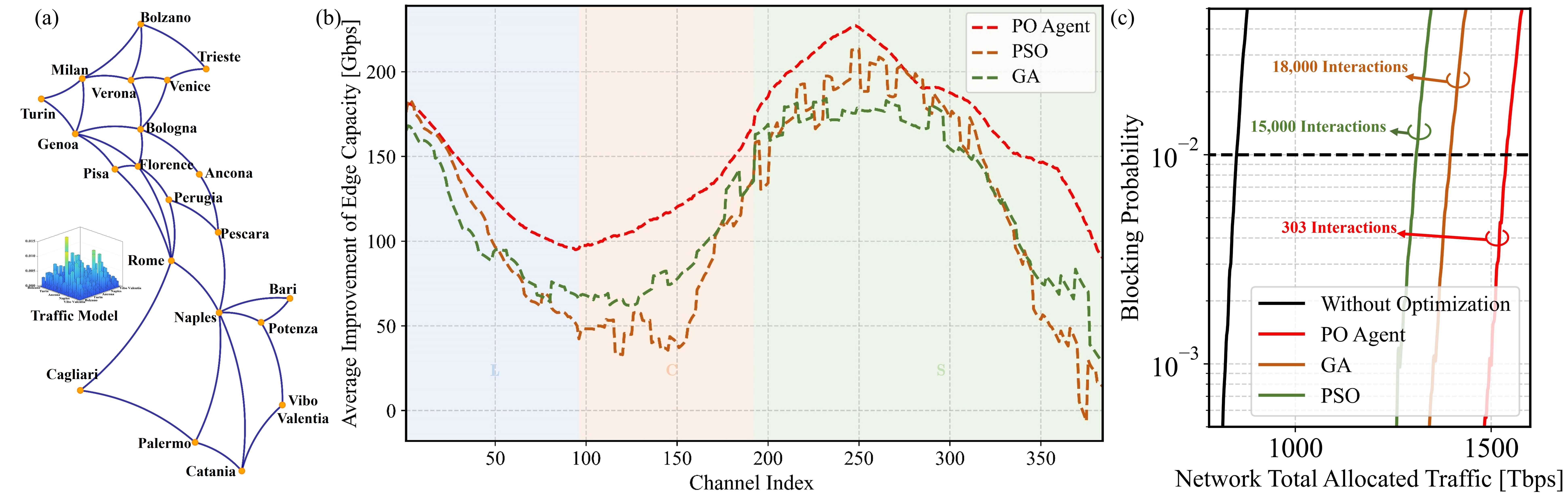}
    \caption{Network-wide scenario and optimization results: (a) Topology of the Italy network alongside the joint probability density function (PDF) of the population-based traffic model; (b) Average improvement in edge capacity across different algorithms; and (c) Total allocated traffic versus blocking probability.}
    \label{fig:fig3}
\end{figure*}

Network power optimization employs a span-by-span strategy utilizing the Local Optimization Global Optimization (LOGO) algorithm \cite{9402935}. Additionally, the launch power and attenuation are optimized sequentially \cite{10214139} to shape the power profile (PP). To evaluate performance under various power optimization methods, we adopt the Statistical Network Assessment Process (SNAP) \cite{7829312}, driven by a population-based non-uniform traffic model. The joint probability density function (JPDF) of this non-uniform model, which determines the frequency of requests between each node pair, is derived from the population statistics \cite{10892225}. We set the Monte Carlo iteration count {\small$N_{mc}$} to 5000 to ensure stable metric convergence. Routing follows a {\small$k$-shortest} path algorithm ({\small$k = 8$}), with First-Fit (FF) wavelength assignment (WA) used during progressive traffic analysis to capture both dynamic and static performance metrics.

For network power optimization, we impose no constraints on the maximum number of interactions for the PO Agent; instead, it determines when to terminate the current optimization process autonomously. In contrast, for PSO and GA, the maximum number of interactions is set to 15,000 and 18,000, and the number of adjacent channels is configured to 10 to enhance optimization performance.

Fig. 3(b) illustrates the average improvement in edge capacity across all channels for every network link. The results demonstrate that the PO Agent achieves the highest average improvement for nearly all channels compared to baseline methods. Fig. 3(c) depicts the relationship between the total allocated traffic enhancement and the blocking probability (BP) threshold, set at 0.01. Near this threshold, the PO Agent delivers the most significant improvement, achieving a total throughput of 689.0 Tbps. In comparison, PSO and GA achieve improvements of 458.1 Tbps and 546.0 Tbps, respectively. Additionally, the PO Agent demonstrates superior efficiency, requiring an average of only 303 interactions per PP to converge.

\section{Conclusion}

In this paper, the PO Agent is specifically designed to optimize link power in multi-band transmission systems, significantly enhancing optimization efficiency. In a span-level study across S+C+L bands, the PO Agent outperforms both PSO and GA across various optimization objectives. Furthermore, in a network-wide evaluation using the real-world Italy topology, the PO Agent achieves a significant enhancement of 689.0 Tbps in total allocated traffic, requiring an average of only 303 interactions per PP.

%-------------------------------------------------- Acknowledgements Section -------------------------------------------------------%
\clearpage
\section{Acknowledgements}

This work was supported in part by National Key R\&D Program of China (2023YFB2905400), National Natural Science Foundation of China 
(62501388), and Shanghai Jiao Tong University 2030 Initiative. 

The first three authors contributed equally to this work.

%-------------------------------------------------- Bibliography Section -------------------------------------------------------%
% see also https://tex.stackexchange.com/questions/55030/text-before-references-but-after-bibliography-title-with-bibtex as of 2024-02-29
\printbibliography

@article{9076329,
  author={Ferrari, Alessio and Napoli, Antonio and Fischer, Johannes K. and Costa, Nelson and D’Amico, Andrea and Pedro, João and Forysiak, Wladek and Pincemin, Erwan and Lord, Andrew and Stavdas, Alexandros and Gimenez, Juan Pedro F.-P. and Roelkens, Gunther and Calabretta, Nicola and Abrate, Silvio and Sommerkorn-Krombholz, Bernd and Curri, Vittorio},
  journal={Journal of Lightwave Technology}, 
  title={Assessment on the Achievable Throughput of Multi-Band ITU-T G.652.D Fiber Transmission Systems}, 
  year={2020},
  volume={38},
  number={16},
  pages={4279-4291},
  doi={10.1109/JLT.2020.2989620}}

@article{10530896,
  author={Escobar-Landero, Salma and Lorences-Riesgo, Abel and Zhao, Xiaohui and Frignac, Yann and Charlet, Gabriel},
  journal={Journal of Lightwave Technology}, 
  title={S+C+L High-Capacity Transmission Systems: Challenges and Opportunities}, 
  year={2024},
  volume={42},
  number={12},
  pages={4260-4270},
  doi={10.1109/JLT.2024.3401540}}

@article{10892225,
  author={Arpanaei, Farhad and Ranjbar Zefreh, Mahdi and Jiang, Yanchao and Poggiolini, Pierluigi and Ghodsifar, Kimia and Beyranvand, Hamzeh and Natalino, Carlos and Monti, Paolo and Napoli, Antonio and Rivas-Moscoso, José M. and González de Dios, Óscar and Fernández-Palacios, Juan Pedro and Dobre, Octavia A. and Hernández, José Alberto and Larrabeiti, David},
  journal={IEEE Journal on Selected Areas in Communications}, 
  title={Synergizing Hyper-Accelerated Power Optimization and Wavelength-Dependent QoT-Aware Cross-Layer Design in Next-Generation Multi-Band EONs}, 
  year={2025},
  volume={43},
  number={5},
  pages={1840-1855},
  doi={10.1109/JSAC.2025.3543528}}

@ARTICLE{9580565,
  author={Rapp, Lutz and Eiselt, Michael},
  journal={Journal of Lightwave Technology}, 
  title={Optical Amplifiers for Multi–Band Optical Transmission Systems}, 
  year={2022},
  volume={40},
  number={6},
  pages={1579-1589},
  doi={10.1109/JLT.2021.3120944}}

@article{9402935,
  author={Correia, Bruno and Sadeghi, Rasoul and Virgillito, Emanuele and Napoli, Antonio and Costa, Nelson and Pedro, Joao and Curri, Vittorio},
  journal={Journal of Optical Communications and Networking}, 
  title={Power control strategies and network performance assessment for C+L+S multiband optical transport}, 
  year={2021},
  volume={13},
  number={7},
  pages={147-157},
  doi={10.1364/JOCN.419293}}

@article{10214139,
  author={Zhang, Yao and Pang, Xuhao and Song, Yuchen and Wang, Yu and Zhou, Ying and Zhu, Hong and Zhang, Lifang and Fan, Yanlin and Guo, Zhibin and Huang, Shanguo and Zhang, Min and Wang, Danshi},
  journal={Journal of Lightwave Technology}, 
  title={Optical Power Control for GSNR Optimization Based on C+L-Band Digital Twin Systems}, 
  year={2024},
  volume={42},
  number={1},
  pages={95-105},
  doi={10.1109/JLT.2023.3303783}}

@ARTICLE{9035688,
  author={Ferrari, Alessio and Filer, Mark and Balasubramanian, Karthikeyan and Yin, Yawei and Le Rouzic, Esther and Kundrat, Jan and Grammel, Gert and Galimberti, Gabriele and Curri, Vittorio},
  journal={Journal of Optical Communications and Networking}, 
  title={GNPy: an open source application for physical layer aware open optical networks}, 
  year={2020},
  volume={12},
  number={6},
  pages={C31-C40},
  doi={10.1364/JOCN.382906}}

@ARTICLE{9082752,
  author={Semrau, Daniel and Sillekens, Eric and Bayvel, Polina and Killey, Robert I.},
  journal={Journal of Optical Communications and Networking}, 
  title={Modeling and mitigation of fiber nonlinearity in wideband optical signal transmission [Invited]}, 
  year={2020},
  volume={12},
  number={6},
  pages={C68-C76},
  doi={10.1364/JOCN.382267}}

@article{Luo:22,
author = {Huaijian Luo and Jianing Lu and Zhuili Huang and Changyuan Yu and Chao Lu},
journal = {Opt. Express},
number = {1},
pages = {664--675},
publisher = {Optica Publishing Group},
title = {Optimization strategy of power control for C$+$L$+$S band transmission using a simulated annealing algorithm},
volume = {30},
month = {Jan},
year = {2022},
url = {https://opg.optica.org/oe/abstract.cfm?URI=oe-30-1-664},
doi = {10.1364/OE.439635},
}

@INPROCEEDINGS{9748510,
  author={Pang, Xuhao and Li, Shengnan and Fan, Qirui and Zhang, Min and Lu, Chao and Lau, Alan Pak Tao and Wang, Danshi},
  booktitle={2022 Optical Fiber Communications Conference and Exhibition (OFC)}, 
  title={Digital Twin-Assisted Optical Power Allocation for Flexible and Customizable SNR Optimization}, 
  year={2022},
  volume={},
  number={},
  pages={1-3},
  doi={}}

@article{11082171,
  author={Ran, Min and Gong, Miao and Yin, Qiang and Luo, Ming and Huang, Tianye and Li, Xiang},
  booktitle={2025 IEEE 34th Wireless and Optical Communications Conference (WOCC)}, 
  title={Optimization Strategy for Power Control in C+L+S Band Transmission Using Particle Swarm Optimization}, 
  year={2025},
  volume={},
  number={},
  pages={319-323},
  doi={10.1109/WOCC63563.2025.11082171}}

@misc{yao2023reactsynergizingreasoningacting,
      title={ReAct: Synergizing Reasoning and Acting in Language Models}, 
      author={Shunyu Yao and Jeffrey Zhao and Dian Yu and Nan Du and Izhak Shafran and Karthik Narasimhan and Yuan Cao},
      year={2023},
      eprint={2210.03629},
      archivePrefix={arXiv},
      primaryClass={cs.CL},
      url={https://arxiv.org/abs/2210.03629}, 
}

@INPROCEEDINGS{9492399,
  author={Shevchenko, Nikita A. and Nallaperuma, Sam and Savory, Seb J.},
  booktitle={2021 International Conference on Optical Network Design and Modeling (ONDM)}, 
  title={Ultra-Wideband Information Throughput Attained via Launch Power Allocation}, 
  year={2021},
  volume={},
  number={},
  pages={1-3},
  doi={10.23919/ONDM51796.2021.9492399}}

@article{langchain,
author = {Topsakal, Oguzhan and Akinci, T. Cetin},
year = {2023},
month = {07},
pages = {1050-1056},
title = {Creating Large Language Model Applications Utilizing LangChain: A Primer on Developing LLM Apps Fast},
volume = {1},
journal = {International Conference on Applied Engineering and Natural Sciences},
doi = {10.59287/icaens.1127}
}

@misc{deepseekai2025deepseekv32pushingfrontieropen,
      title={DeepSeek-V3.2: Pushing the Frontier of Open Large Language Models}, 
      author={DeepSeek-AI and Aixin Liu and Aoxue Mei and Bangcai Lin and Bing Xue and Bingxuan Wang and Bingzheng Xu and Bochao Wu and Bowei Zhang and Chaofan Lin and Chen Dong and Chengda Lu and Chenggang Zhao and Chengqi Deng and Chenhao Xu and Chong Ruan and Damai Dai and Daya Guo and Dejian Yang and Deli Chen and Erhang Li and Fangqi Zhou and Fangyun Lin and Fucong Dai and Guangbo Hao and Guanting Chen and Guowei Li and H. Zhang and Hanwei Xu and Hao Li and Haofen Liang and Haoran Wei and Haowei Zhang and Haowen Luo and Haozhe Ji and Honghui Ding and Hongxuan Tang and Huanqi Cao and Huazuo Gao and Hui Qu and Hui Zeng and Jialiang Huang and Jiashi Li and Jiaxin Xu and Jiewen Hu and Jingchang Chen and Jingting Xiang and Jingyang Yuan and Jingyuan Cheng and Jinhua Zhu and Jun Ran and Junguang Jiang and Junjie Qiu and Junlong Li and Junxiao Song and Kai Dong and Kaige Gao and Kang Guan and Kexin Huang and Kexing Zhou and Kezhao Huang and Kuai Yu and Lean Wang and Lecong Zhang and Lei Wang and Liang Zhao and Liangsheng Yin and Lihua Guo and Lingxiao Luo and Linwang Ma and Litong Wang and Liyue Zhang and M. S. Di and M. Y Xu and Mingchuan Zhang and Minghua Zhang and Minghui Tang and Mingxu Zhou and Panpan Huang and Peixin Cong and Peiyi Wang and Qiancheng Wang and Qihao Zhu and Qingyang Li and Qinyu Chen and Qiushi Du and Ruiling Xu and Ruiqi Ge and Ruisong Zhang and Ruizhe Pan and Runji Wang and Runqiu Yin and Runxin Xu and Ruomeng Shen and Ruoyu Zhang and S. H. Liu and Shanghao Lu and Shangyan Zhou and Shanhuang Chen and Shaofei Cai and Shaoyuan Chen and Shengding Hu and Shengyu Liu and Shiqiang Hu and Shirong Ma and Shiyu Wang and Shuiping Yu and Shunfeng Zhou and Shuting Pan and Songyang Zhou and Tao Ni and Tao Yun and Tian Pei and Tian Ye and Tianyuan Yue and Wangding Zeng and Wen Liu and Wenfeng Liang and Wenjie Pang and Wenjing Luo and Wenjun Gao and Wentao Zhang and Xi Gao and Xiangwen Wang and Xiao Bi and Xiaodong Liu and Xiaohan Wang and Xiaokang Chen and Xiaokang Zhang and Xiaotao Nie and Xin Cheng and Xin Liu and Xin Xie and Xingchao Liu and Xingkai Yu and Xingyou Li and Xinyu Yang and Xinyuan Li and Xu Chen and Xuecheng Su and Xuehai Pan and Xuheng Lin and Xuwei Fu and Y. Q. Wang and Yang Zhang and Yanhong Xu and Yanru Ma and Yao Li and Yao Li and Yao Zhao and Yaofeng Sun and Yaohui Wang and Yi Qian and Yi Yu and Yichao Zhang and Yifan Ding and Yifan Shi and Yiliang Xiong and Ying He and Ying Zhou and Yinmin Zhong and Yishi Piao and Yisong Wang and Yixiao Chen and Yixuan Tan and Yixuan Wei and Yiyang Ma and Yiyuan Liu and Yonglun Yang and Yongqiang Guo and Yongtong Wu and Yu Wu and Yuan Cheng and Yuan Ou and Yuanfan Xu and Yuduan Wang and Yue Gong and Yuhan Wu and Yuheng Zou and Yukun Li and Yunfan Xiong and Yuxiang Luo and Yuxiang You and Yuxuan Liu and Yuyang Zhou and Z. F. Wu and Z. Z. Ren and Zehua Zhao and Zehui Ren and Zhangli Sha and Zhe Fu and Zhean Xu and Zhenda Xie and Zhengyan Zhang and Zhewen Hao and Zhibin Gou and Zhicheng Ma and Zhigang Yan and Zhihong Shao and Zhixian Huang and Zhiyu Wu and Zhuoshu Li and Zhuping Zhang and Zian Xu and Zihao Wang and Zihui Gu and Zijia Zhu and Zilin Li and Zipeng Zhang and Ziwei Xie and Ziyi Gao and Zizheng Pan and Zongqing Yao and Bei Feng and Hui Li and J. L. Cai and Jiaqi Ni and Lei Xu and Meng Li and Ning Tian and R. J. Chen and R. L. Jin and S. S. Li and Shuang Zhou and Tianyu Sun and X. Q. Li and Xiangyue Jin and Xiaojin Shen and Xiaosha Chen and Xinnan Song and Xinyi Zhou and Y. X. Zhu and Yanping Huang and Yaohui Li and Yi Zheng and Yuchen Zhu and Yunxian Ma and Zhen Huang and Zhipeng Xu and Zhongyu Zhang and Dongjie Ji and Jian Liang and Jianzhong Guo and Jin Chen and Leyi Xia and Miaojun Wang and Mingming Li and Peng Zhang and Ruyi Chen and Shangmian Sun and Shaoqing Wu and Shengfeng Ye and T. Wang and W. L. Xiao and Wei An and Xianzu Wang and Xiaowen Sun and Xiaoxiang Wang and Ying Tang and Yukun Zha and Zekai Zhang and Zhe Ju and Zhen Zhang and Zihua Qu},
      year={2025},
      eprint={2512.02556},
      archivePrefix={arXiv},
      primaryClass={cs.CL},
      url={https://arxiv.org/abs/2512.02556}, 
}

@inproceedings{ijcai2024p890,
  title     = {Large Language Model Based Multi-agents: A Survey of Progress and Challenges},
  author    = {Guo, Taicheng and Chen, Xiuying and Wang, Yaqi and Chang, Ruidi and Pei, Shichao and Chawla, Nitesh V. and Wiest, Olaf and Zhang, Xiangliang},
  booktitle = {Proceedings of the Thirty-Third International Joint Conference on
               Artificial Intelligence, {IJCAI-24}},
  publisher = {International Joint Conferences on Artificial Intelligence Organization},
  editor    = {Kate Larson},
  pages     = {8048--8057},
  year      = {2024},
  month     = {8},
  note      = {Survey Track},
  doi       = {10.24963/ijcai.2024/890},
  url       = {https://doi.org/10.24963/ijcai.2024/890},
}

@inproceedings{li-etal-2025-exposing,
    title = "Exposing Numeracy Gaps: A Benchmark to Evaluate Fundamental Numerical Abilities in Large Language Models",
    author = "Li, Haoyang  and
      Chen, Xuejia  and
      Xu, Zhanchao  and
      Li, Darian  and
      Hu, Nicole  and
      Teng, Fei  and
      Li, Yiming  and
      Qiu, Luyu  and
      Zhang, Chen Jason  and
      Qing, Li  and
      Chen, Lei",
    editor = "Che, Wanxiang  and
      Nabende, Joyce  and
      Shutova, Ekaterina  and
      Pilehvar, Mohammad Taher",
    booktitle = "Findings of the Association for Computational Linguistics: ACL 2025",
    month = jul,
    year = "2025",
    address = "Vienna, Austria",
    publisher = "Association for Computational Linguistics",
    url = "https://aclanthology.org/2025.findings-acl.1026/",
    doi = "10.18653/v1/2025.findings-acl.1026",
    pages = "20004--20026",
    ISBN = "979-8-89176-256-5"
}

@ARTICLE{7829312,
  author={Curri, Vittorio and Cantono, Mattia and Gaudino, Roberto},
  journal={Journal of Lightwave Technology}, 
  title={Elastic All-Optical Networks: A New Paradigm Enabled by the Physical Layer. How to Optimize Network Performances?}, 
  year={2017},
  volume={35},
  number={6},
  pages={1211-1221},
  doi={10.1109/JLT.2017.2657231}}

@ARTICLE{10789615,
  author={Matzner, Robin and Ahuja, Akanksha and Sadeghi, Rasoul and Doherty, Michael and Beghelli, Alejandra and Savory, Seb J. and Bayvel, Polina},
  journal={Journal of Optical Communications and Networking}, 
  title={Topology Bench: systematic graph-based benchmarking for core optical networks}, 
  year={2025},
  volume={17},
  number={1},
  pages={7-27},
  doi={10.1364/JOCN.534477}}

@ARTICLE{10187397,
  author={Zhuge, Qunbi and Liu, Xiaomin and Zhang, Yihao and Cai, Meng and Liu, Yichen and Qiu, Qizhi and Zhong, Xueying and Wu, Jiaping and Gao, Ruoxuan and Yi, Lilin and Hu, Weisheng},
  journal={Journal of Optical Communications and Networking}, 
  title={Building a digital twin for intelligent optical networks [Invited Tutorial]}, 
  year={2023},
  volume={15},
  number={8},
  pages={C242-C262},
  doi={10.1364/JOCN.483600}}

@ARTICLE{9960788,
  author={Song, Yuchen and Fan, Qirui and Lu, Chao and Wang, Danshi and Lau, Alan Pak Tao},
  journal={Journal of Lightwave Technology}, 
  title={Efficient Three-Step Amplifier Configuration Algorithm for Dynamic C+L-Band Links in Presence of Stimulated Raman Scattering}, 
  year={2023},
  volume={41},
  number={5},
  pages={1445-1453},
  doi={10.1109/JLT.2022.3223919}}

@inproceedings{
liu2025agenthpo,
title={Agent{HPO}: Large Language Model Agent for  Hyper-Parameter Optimization},
author={Siyi Liu and Chen Gao and Yong Li},
booktitle={The Second Conference on Parsimony and Learning (Proceedings Track)},
year={2025},
url={https://openreview.net/forum?id=HU3yfXcoKU}
}

@misc{xi2023risepotentiallargelanguage,
      title={The Rise and Potential of Large Language Model Based Agents: A Survey}, 
      author={Zhiheng Xi and Wenxiang Chen and Xin Guo and Wei He and Yiwen Ding and Boyang Hong and Ming Zhang and Junzhe Wang and Senjie Jin and Enyu Zhou and Rui Zheng and Xiaoran Fan and Xiao Wang and Limao Xiong and Yuhao Zhou and Weiran Wang and Changhao Jiang and Yicheng Zou and Xiangyang Liu and Zhangyue Yin and Shihan Dou and Rongxiang Weng and Wensen Cheng and Qi Zhang and Wenjuan Qin and Yongyan Zheng and Xipeng Qiu and Xuanjing Huang and Tao Gui},
      year={2023},
      eprint={2309.07864},
      archivePrefix={arXiv},
      primaryClass={cs.AI},
      url={https://arxiv.org/abs/2309.07864}, 
}

@ARTICLE{7470251,
  author={Roberts, Ian and Kahn, Joseph M. and Boertjes, David},
  journal={Journal of Lightwave Technology}, 
  title={Convex Channel Power Optimization in Nonlinear WDM Systems Using Gaussian Noise Model}, 
  year={2016},
  volume={34},
  number={13},
  pages={3212-3222},
  doi={10.1109/JLT.2016.2569073}}

\vspace{-4mm}

%%%%%%%%%%%%%%%%%%%%%%%%%%%%%%%%%%%%%%%%%%%%%
%---------------------------------------------- End of Document -----------------------------------------------%
\end{document}